\documentclass[11pt]{article}
\usepackage{epsfig}
\usepackage{amsfonts}
\usepackage{amsmath}
\usepackage{bbm,bm}
\usepackage{cite}
\allowdisplaybreaks[1]

\newcommand{\nr}[1]{(\ref{#1})}
\newcommand{\tr}{{\rm Tr\,}}

\newcommand{\fr}[2]{{\frac{#1}{#2}}}

\newcommand{\msbar}{{\overline{\mbox{\rm MS}}}}

\renewcommand{\vec}[1]{{\bf #1}}
\newcommand{\la}[1]{\label{#1}}
\newcommand{\be}{\begin{equation}}
\newcommand{\ee}{\end{equation}}
\newcommand{\ba}{\begin{eqnarray}}
\newcommand{\ea}{\end{eqnarray}}
\newcommand{\rmi}[1]{{\mbox{\scriptsize #1}}}

\newcommand{\eq}{eq.~}
\newcommand{\eqs}{eqs.~}
\newcommand{\se}{sec.~}

\newcommand{\fig}{fig.~}
\newcommand{\figs}{figs.~}
\newcommand{\tinymsbar}{{\overline{\mbox{\tiny\rm{MS}}}}}
\newcommand{\Lambdamsbar}{{\Lambda_\tinymsbar}}

\newcommand{\alphas}{\alpha_{\rm s}}
\newcommand{\Nf}{N_{\rm f}}
\newcommand{\Nc}{N_{\rm c}}
\newcommand{\Ns}{N_{\rm s}}
\newcommand{\Tc}{T_{\rm c}}
\newcommand{\rO}{r^{ }_0}
\newcommand{\tO}{t^{ }_0}

\newcommand{\Nt}{N^{ }_\tau}
\newcommand{\CF}{C_\rmii{F}}

\newcommand{\rmO}{{\mathcal{O}}}
\newcommand{\bmu}{\bar\mu}

\def\lsi{\raise0.3ex\hbox{$<$\kern-0.75em\raise-1.1ex\hbox{$\sim$}}}
\def\gsi{\raise0.3ex\hbox{$>$\kern-0.75em\raise-1.1ex\hbox{$\sim$}}}
\newcommand{\lsim}{\mathop{\lsi}}
\newcommand{\gsim}{\mathop{\gsi}}

\newcommand{\rmii}[1]{{\mbox{\tiny\rm{#1}}}}

\newcommand{\re}{\mathop{\mbox{Re}}}

\newcommand{\Tint}[1]{{\hbox{$\sum$}\!\!\!\!\!\!\!\int\,}_{\!\!\!\!\raise-0.9ex\hbox{$\scriptstyle{#1}$}}}
\newcommand{\Tinti}[1]{{{\Sigma}\!\!\!\!\raise0.3ex\hbox{$\int$}_\rmii{${#1}$}}}

\newcommand{\bi}{\begin{itemize}}
\newcommand{\ei}{\end{itemize}}
\newcommand{\hide}[1]{ }

\makeatletter \@addtoreset{equation}{section} \makeatother

\makeatletter
\renewcommand\section{\@startsection {section}{1}{\z@}%
                                   {-5.5ex \@plus -1ex \@minus -.2ex}% bfr-
                                   {2.3ex \@plus.2ex}%
                                   {\normalfont\large\bfseries}}
\renewcommand\subsection{\@startsection{subsection}{2}{\z@}%
                                     {-3.25ex\@plus -1ex \@minus -.2ex}%
                                     {1.5ex \@plus .2ex}%
                                     {\normalfont\normalsize\bfseries}}
\renewcommand\thesection {\@arabic\c@section}
\renewcommand\thesubsection   {\thesection.\@arabic\c@subsection}
\renewcommand{\@seccntformat}[1]{%
\csname the#1\endcsname.\hspace{1.0em}}
\makeatother
\begin{document}

\flushbottom

\begin{titlepage}

\begin{flushright}
BI-TP 2015/11 \\  
HIP-2015-28/TH \\ % Notes M.L. \\ 
October 2015
\end{flushright}
\begin{centering}

\vspace*{4mm}

{\Large{\bf
 A non-perturbative estimate of the heavy quark
 \\[2mm] momentum diffusion coefficient
}} 

\vspace{0.8cm}

A.~Francis$^{\rm a}$, 
O.~Kaczmarek$^{\rm b}$, 
M.~Laine$^{\rm c,d}$, 
T.~Neuhaus$^{\rm e,}$\footnote{%
 In memoriam (27 October 1956 -- 1 June 2015).
 } 
 and 
H.~Ohno$^{\rm f,g}$

\vspace{0.8cm}

$^\rmi{a}$%
{\em
        Dept.\ of Physics and Astronomy, 
        York University, % 4700 Keele St., 
        Toronto, ON M3J1P3, Canada\\
}

\vspace{0.3cm}

$^\rmi{b}$%
{\em
       Faculty of Physics, University of Bielefeld, 
        33501 Bielefeld, Germany\\
}

\vspace{0.3cm}

$^\rmi{c}$%
{\em
 ITP, AEC, University of Bern,  
 Sidlerstrasse 5, 3012 Bern, Switzerland\\
}

\vspace*{0.3cm}

$^\rmi{d}$%
{\em
 Helsinki Institute of Physics, 
 P.O.Box 64, 
 00014 University of Helsinki, Finland\\
}

\vspace*{0.3cm}

$^\rmi{e}$%
{\em
       Institute for Advanced Simulation,  
       FZ J\"ulich, % Supercomputing Centre, \\
       52425 J\"ulich, Germany\\
}

\vspace*{0.3cm}

$^\rmi{f}$%
{\em
        Center for Computational Sciences, University of Tsukuba, 
        % Tsukuba, 
        Ibaraki 305-8577, Japan\\
}

\vspace*{0.3cm}

$^\rmi{g}$%
{\em
        Physics Department, Brookhaven National Laboratory, 
        Upton, NY 11973, USA\\
}

%        \email{francis@kph.uni-mainz.de}, 
%        \email{okacz@physik.uni-bielefeld.de}, 
%        \email{laine@itp.unibe.ch},
%        \email{mmueller@physik.uni-bielefeld.de},
%        \email{hono@quark.phy.bnl.gov}

\vspace*{0.8cm}

\mbox{\bf Abstract}
 
\end{centering}

\vspace*{0.3cm}
 
\noindent
We estimate the momentum diffusion coefficient of a heavy quark within 
a pure SU(3) plasma at a temperature of about $1.5\Tc$. Large-scale Monte 
Carlo simulations on a series of lattices extending up to $192^3 \times 48$ 
permit us to carry out a continuum extrapolation of the so-called 
colour-electric imaginary-time correlator. The extrapolated correlator 
is analyzed with the help of theoretically motivated models for the 
corresponding spectral function. Evidence for a non-zero 
transport coefficient is found and, incorporating systematic uncertainties 
reflecting model assumptions, we obtain $\kappa = (1.8 - 3.4)\, T^3$. 
This implies 
that the ``drag coefficient'', characterizing the time scale at which 
heavy quarks adjust to hydrodynamic flow, is 
$\eta^{-1}_\rmii{$D$} = (1.8 - 3.4) (\Tc/T)^2 (M/\mbox{1.5GeV})\,$fm/c, 
where $M$ is the heavy quark kinetic mass. The results apply to bottom
and, with somewhat larger systematic uncertainties, to charm quarks.   

\vfill

%% %\noindent
%% %PACS numbers: 
%% %11.10.Wx, %        Finite temperature field theory
%% { %11.15.Ha, %        Lattice gauge theory } 
%% %12.38.Bx, %        Perturbative calculations in QCD
%% %12.38.Mh, %        Quark--gluon plasma
%% 12.39.Hg   Heavy quark effective theory 
%% %14.40.Nd, %        Bottom mesons
%% %\\
%% %Keywords: Thermal Field Theory, Neutrino Physics, Resummation
 
%% \vspace*{1cm}
  
\vfill

\end{titlepage}

%%%%%%%%%%%%%%%%%%%%%%%%%%%%% SECTION %%%%%%%%%%%%%%%%%%%%%%%%%%%%%%%%%%%%
%
\section{Introduction}

Within linear response theory the rate at which a system relaxes
towards local thermal equilibrium is characterized by quantities known 
as transport coefficients. Different transport
coefficients parametrize different types of perturbations. 
If we focus on a conserved particle number, such as a quark 
flavour,\footnote{%
 Weak interactions play no role within the lifetime $\lsim 20$~fm/c
 of a fireball generated in a heavy ion collision. 
 }
which is initially 
distributed unevenly, such as in a broad jet cone, then the relevant
transport coefficient is the diffusion coefficient. In QCD, 
there is a separate diffusion coefficient related to light flavours, 
and to heavy flavours such as charm and bottom. A closely related
quantity is the electrical conductivity, which can be expressed
as a weighted sum over the flavour diffusion coefficients. 

The determination of transport coefficients related to strong interactions
at temperatures of a few hundred MeV is an important goal for lattice 
QCD. It is a challenging task, 
given that lattice QCD is formulated in a Euclidean spacetime whereas 
transport coefficients are real-time 
quantities (for a review, see ref.~\cite{hbm}). Nevertheless, motivated
by the ongoing heavy ion collision program, large-scale efforts have
been undertaken. For example, for the light-quark contribution to  
electrical conductivity, recent works can be found
in refs.~\cite{cond1,cond2a,cond3,cond4,cond5,cond2b}. 
 
The focus of the present study is the diffusion coefficient associated
with heavy quarks. It has been one of the major qualitative discoveries
of the heavy ion collision program at RHIC and LHC that charm quarks
appear to flow about as efficiently as light quarks do 
(see, e.g., refs.~\cite{meas1,meas2,meas3} and references therein). 
That a flow develops is an example of a relaxation towards
local thermal equilibrium. 
It is then a theoretical challenge to explain this behaviour from 
the laws of QCD~\cite{mt}. A next-to-leading order (NLO)
computation in perturbation
theory indicates the presence of a large correction
towards strong interactions~\cite{chm1}, 
and strong interactions have also been observed in 
$\mathcal{N} = 4$ Super-Yang-Mills 
theory~\cite{ads1,ads2,ct,chm2}. Furthermore classical lattice 
gauge theory simulations~\cite{mink} and analyses in the 
confined phase~\cite{conf1,conf3,conf5,conf4,conf2}
are consistent with strong interactions, 
and various other approaches are being pursued 
in the same vein~\cite{model1,model3,model2,model5,model7,model4,model6}  
(for a review, see ref.~\cite{rev}). Heavy quark diffusion
also happens to pose an ideal ground for more general theoretical 
investigations of non-equilibrium thermodynamics~\cite{ya3}. 
In any case, ultimately the problem
needs to be addressed with lattice simulations. 

For $M \gg \pi T$, where $M$ is the heavy-quark ``kinetic mass'',
the lattice determination of the heavy quark diffusion coefficient
can be reduced to a purely gluonic
measurement~\cite{eucl}.  Here we report the final results 
of a multiyear study of the relevant observable in the deconfined
phase of pure SU(3) gauge theory. 
It has been demonstrated a while ago that, 
with advanced numerical methods, a signal can be obtained 
at a fixed lattice spacing~\cite{latt_a,kappaE,latt_c}. However
the issues of renormalization, taking the continuum limit, and
analytic continuation had not been brought into conclusion. 
Even though further improvements are needed and 
can be foreseen on all of these fronts, 
the purpose of the current paper is to present an analysis which
offers a ``minimal'' practical answer to the main open points. 

The plan of this paper is the following. 
After introducing the observable and reviewing the techniques
that have been used for its determination
and the data sets that have been collected
(\se\ref{se:meas}), we carry out a continuum extrapolation 
in \se\ref{se:continuum}. The simple structure of 
the associated spectral function, as revealed by previous theoretical
works, allows us to attack the difficult problem of analytic 
continuation through tightly constrained models, as well as a variant
of the so-called Backus-Gilbert method (\se\ref{se:spectral}).
Our final results and some future prospects are presented
in \se\ref{se:concl}.

%%%%%%%%%%%%%%%%%%%%%%%%%%%%% SECTION %%%%%%%%%%%%%%%%%%%%%%%%%%%%%%%%%%%%
%
\section{Measurements}
\la{se:meas}

%%%%%%%%%%%%%%%%%%%%% TABLE %%%%%%%%%%%%%%%%%%%%%%%%%%%%%%%%%%%%%
%
\begin{table}[t]

\small{
\begin{center}
\begin{tabular}{lllllllll}
 \hline \\[-3mm]
 $\beta^{ }_0$ &
 $\Ns^3 \times \Nt$ &
 confs & 
 $ T \sqrt{\tO}{ }^{\rmi{(imp)}} $ & 
 $ \left. T / \Tc \right|^\rmi{(imp)}_{\tO} $  & 
 $ T \sqrt{\tO}{ }^{\rmi{(clov)}} $ & 
 $ \left. T / \Tc \right|^\rmi{(clov)}_{\tO} $  & 
 $ T \rO $ & 
 $ \left. T / \Tc \right|^{ }_{\rO} $  \\[3mm]
 \hline 
  6.872  & $64^3 \times 16$ & 172  & 0.3770 & 1.52 & 0.3805 &  1.53
    & 1.116 & 1.50 \\ 
  7.035  & $80^3 \times 20$ & 180 & 0.3693 & 1.48 & 0.3739 &  1.50
    & 1.086 & 1.46 \\ 
  7.192  & $96^3 \times 24$ & 160 & 0.3728 & 1.50 & 0.3790 &  1.52
    & 1.089 & 1.46 \\ 
  7.544  & $144^3 \times 36$ & 693 & 0.3791 & 1.52 & 0.3896 & 1.57
    & 1.089 & 1.46 \\ 
  7.793  & $192^3 \times 48$ & 223 & 0.3816 & 1.53 & 0.3955 & 1.59
    & 1.084 & 1.45 \\ 
 \hline 
\end{tabular} 
\end{center}
}

\vspace*{3mm}

\caption[a]{\small
  The lattices included in the current analysis. 
  Conversions to units of $\tO$ are based 
  on ref.~\cite{betac} for the improved Wilson discretization (``imp'');
  on refs.~\cite{betac,flowQCD} for the clover discretization (``clov'')
  of the observable defining $\tO$~\cite{t0}; 
  and on ref.~\cite{betac} for $\rO$~\cite{r0}. 
  Note however that the clover case is not well represented by our
  ansatz~\cite{betac}: $\chi^2$/d.o.f.\ $\approx 42$.
  Conversions to $\Tc$ are based on ref.~\cite{betac}. 
 }
\label{table:params}
\end{table}
%
%%%%%%%%%%%%%%%%%%%%%%%%%%%%%%%%%%%%%%%%%%%%%%%%%%%%%%%%%%%%%%%%%%%%%

%%%%%%%%%%%%%%%%%%%%%%%%%%%%%%% FIGURE %%%%%%%%%%%%%%%%%%%%%%%%%%%%%%%%
%
\begin{figure}[t]

\vspace*{3mm}

\begin{center}
  \begin{minipage}[t]{7.cm}
    \includegraphics*[width=7.cm]{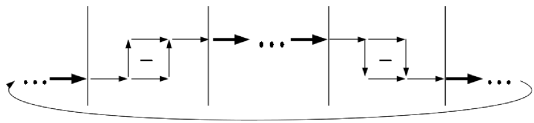}
  \end{minipage}
\end{center}

\vspace*{-3mm}

\caption[a]{\small
 One possible discretization of \eq\nr{GE_final}~\cite{eucl}.
 Different techniques have been used for improving on 
 the statistical signal originating from the links denoted 
 with thick lines, and from the electric field insertions delineated
 with the vertical lines 
 (cf.\ the text).
 }
\label{fig:multilevel-linkintegrated}
\end{figure}
%
%%%%%%%%%%%%%%%%%%%%%%%%%%%%%%%%%%%%%%%%%%%%%%%%%%%%%%%%%%%%%

Using Heavy Quark Effective Theory, 
the force felt by a heavy quark
as it propagates through a gluon plasma can be related
to a ``colour-electric correlator''~\cite{eucl}
(cf.\ also ref.~\cite{ct}),  
\be
 G_\rmii{E}(\tau) \equiv - \fr13 \sum_{i=1}^3 
 \frac{
  \Bigl\langle
   \re\tr \Bigl[
      U(\beta;\tau) \, gE_i(\tau,\vec{0}) \, U(\tau;0) \, gE_i(0,\vec{0})
   \Bigr] 
  \Bigr\rangle
 }{
 \Bigl\langle
   \re\tr [U(\beta;0)] 
 \Bigr\rangle
 }
 \;, \quad \beta \equiv \fr{1}{T}
 \;, \la{GE_final}
\ee
where $gE_i$ denotes the  
colour-electric field, $T$ the temperature, and $U(\tau_2;\tau_1)$
a Wilson line in the Euclidean time direction. 
The discretization of this correlator is not unique; 
we employ the proposal of ref.~\cite{eucl}, 
illustrated in \fig\ref{fig:multilevel-linkintegrated}.

We have measured the discretized correlator in quenched lattice QCD, 
employing the standard Wilson gauge action,  
at a temperature corresponding to about $1.5\Tc$ (the parameters of
the simulations are shown in table~\ref{table:params}). In order to 
obtain a signal for our observable, advanced statistical error
reduction techniques are required. As has been discussed in more
detail in refs.~\cite{kappaE,lat13}, 
we have employed 1000 additional multilevel updates~\cite{lw,shear} 
for the electric field insertions, and link integration~\cite{ppr,fr}
for the straight lines between them. Moreover, in order to 
reduce discretization effects, the imaginary-time separations 
are tree-level improved~\cite{kappaE} in analogy with the 
procedure previously used in other contexts~\cite{r0,shear}.

%%%%%%%%%%%%%%%%%%%%%%%%%%%%%%%%% FIGURE %%%%%%%%%%%%%%%%%%%%%%%%%%%%%%%%%
\begin{figure}[t]

\hspace*{0.4cm}
\centerline{%
 \epsfysize=7.5cm\epsfbox{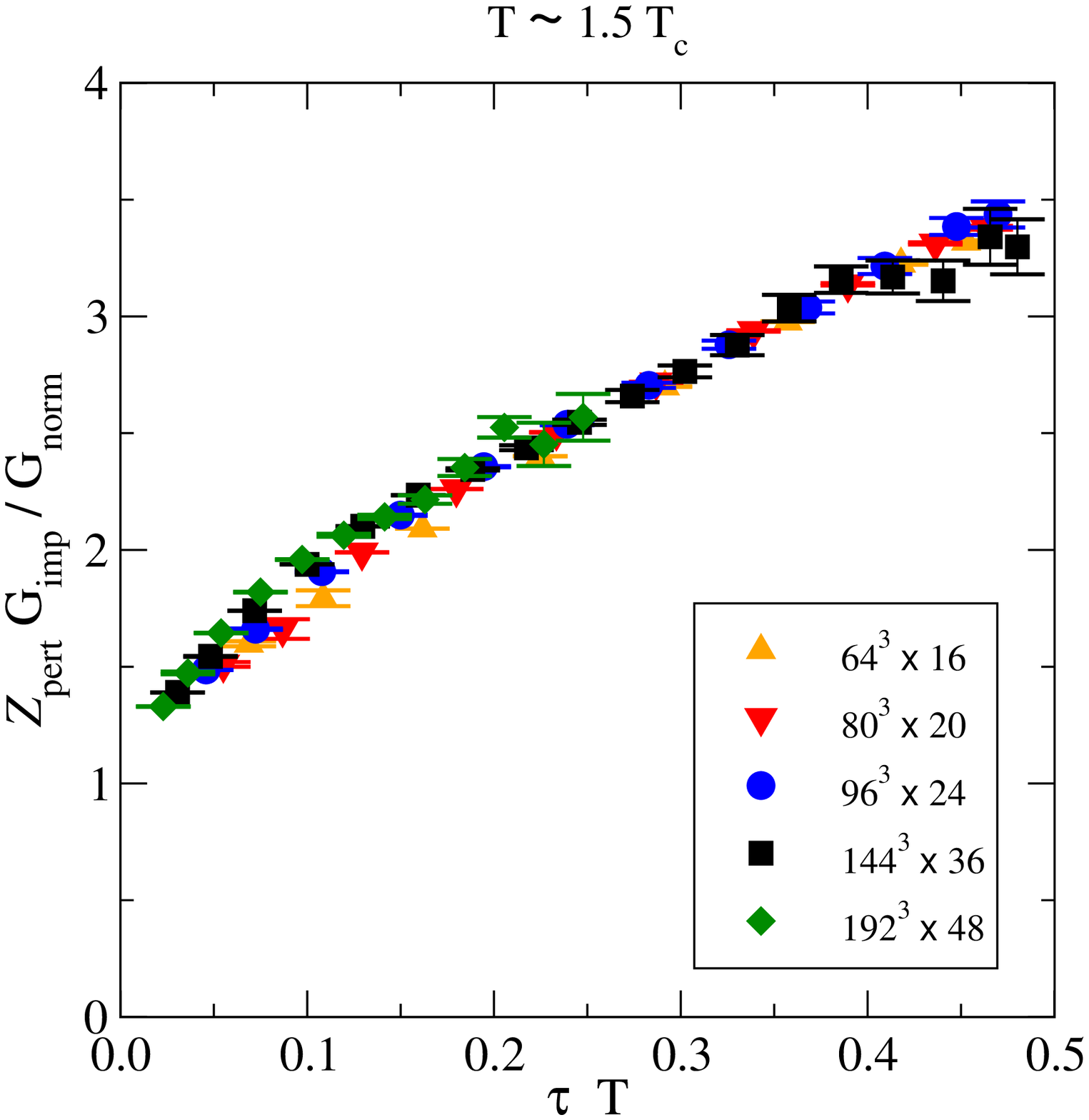}%
 \hspace{0.1cm}
 \epsfysize=7.5cm\epsfbox{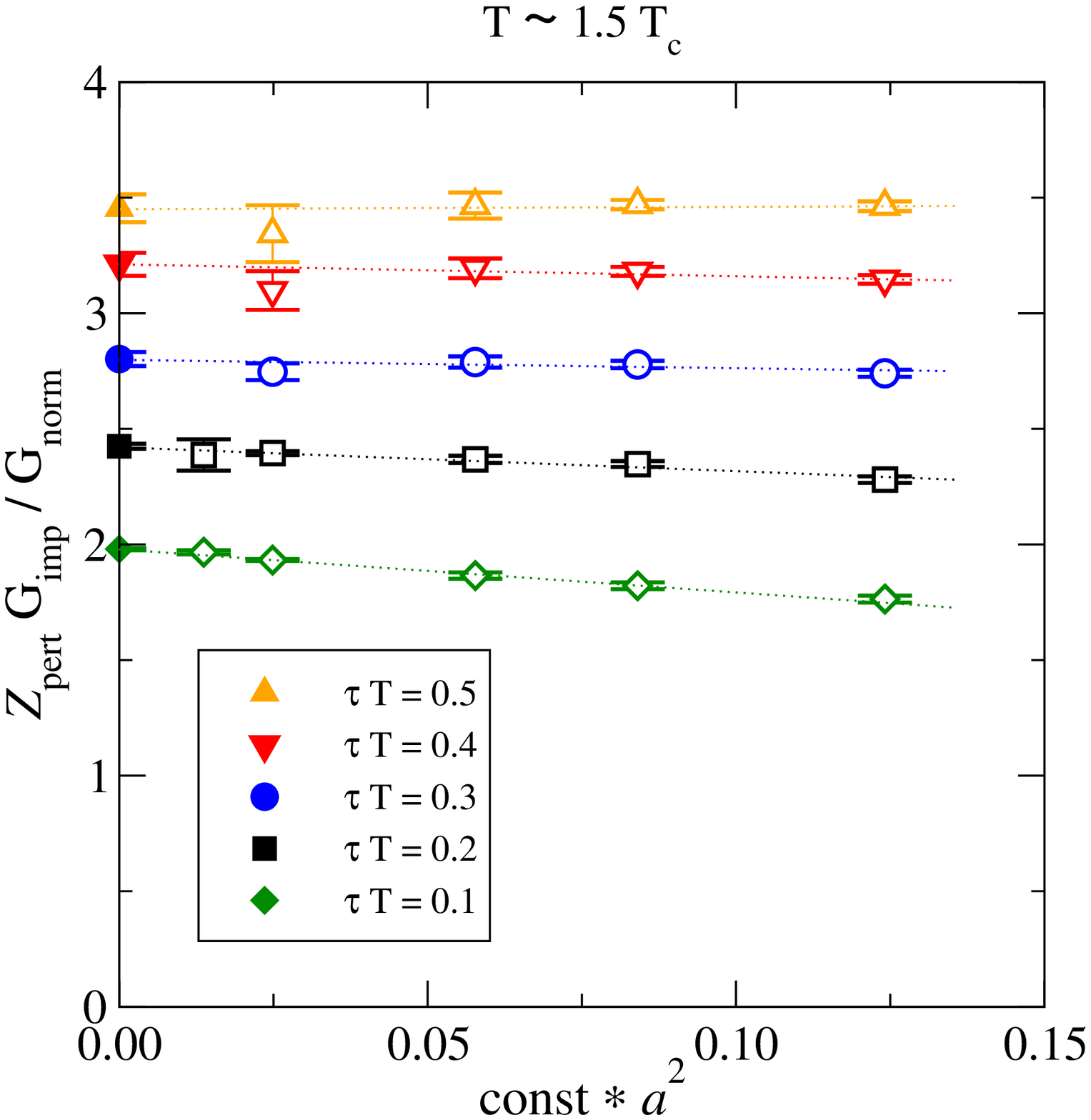}
}

\caption[a]{\small
 Left: Lattice data after perturbative renormalization
 and the use of tree-level improved distances.  
 Right: Illustrations of continuum extrapolations at 
 selected values of $\tau T$ (our final errors are based on 
 a jackknife analysis and differ somewhat from those shown here.)
 % 
 % these correspond to thomas_apr15/gcorr.res
 % 
}

\la{fig:bare}
\end{figure}
%%%%%%%%%%%%%%%%%%%%%%%%%%%%%%%%%%%%%%%%%%%%%%%%%%%%%%%%%%%%%%%%%%%%%%%%%%%

The resulting correlators are illustrated in \fig\ref{fig:bare}(left). 
For better visibility, the correlators have been normalized 
to (cf.\ ref.~\cite{eucl}) 
\be
 G^{ }_\rmi{norm}(\tau)
 \; \equiv \; 
 \pi^2 T^4 \biggl[
 \frac{\cos^2(\pi \tau T)}{\sin^4(\pi \tau T)}
 +\frac{1}{3\sin^2(\pi \tau T)} \biggr] 
 \;, \la{Gnorm}
\ee
which diverges as $1/(\pi^2\tau^4)$ at $\tau\ll\beta$.
The correlators have also been multiplied by a renormalization factor 
as will be explained in \se\ref{se:continuum}. 
As discussed in more detail in 
ref.~\cite{kappaE}, volume dependence lies below
statistical uncertainties in our measurements. In the following, 
we therefore consider a fixed spatial extent in units of 
the temperature, $L^{ }_\rmi{s}= 4/T$. 

%%%%%%%%%%%%%%%%%%%%%%%%%%%%% SECTION %%%%%%%%%%%%%%%%%%%%%%%%%%%%%%%%%%%%
%
\section{Continuum extrapolation}
\la{se:continuum}

Given the lattice data at tree-level improved distances, 
we carry out an extrapolation to the continuum limit. For this
the lattice-measured correlator needs to be multiplied by a 
renormalization factor: 
\ba
 G^{ }_\rmii{E,cont}(\tau) & \equiv &  \mathcal{Z}^{ }_\rmii{E} 
 \, G^{ }_\rmii{E,latt}(\tau) 
 \;. \la{renorm}
\ea
A 1-loop perturbative computation yields~\cite{cc}
\ba
 \mathcal{Z}^{ }_\rmii{E,pert} & = & 
 1 + 0.079 \times  \frac{6}{\beta^{ }_0}  + 
 \rmO\Bigl( 0.079 \times \frac{6}{\beta^{ }_0} \Bigr)^2
 \;, \la{ZE}
\ea
where $\beta^{ }_0 \equiv 6/g_0^2$ is the coupling of the plaquette
term in the Wilson action (cf.\ table~\ref{table:params}). 
Ultimately renormalization should be 
carried out non-perturbatively, however the small coefficient
of the 1-loop term in \eq\nr{ZE} suggests that perturbative
renormalization should yield a reasonable approximation.

Measured results at 4 or 5 different values of $\Nt$ 
(depending on $\tau T$, cf.\ \fig\ref{fig:bare}) are 
interpolated to the values of $\tau T$ that 
are shown in table~\ref{table:taudep}.
At fixed $\tau T$ we extrapolate 
the correlator in $a^2$ to the continuum limit. 
The procedure is illustrated in
\fig\ref{fig:bare}(right) for selected values of $\tau T$. 
The resulting continuum limit 
is shown in \fig\ref{fig:taudep}(left), and the results
are tabulated in table~\ref{table:taudep}. 
 The results and errors of continuum-extrapolated correlation functions 
 were obtained from a combined jackknife analysis. A covariance matrix
 for the continuum correlator was also estimated within this
 analysis (however the errors are strongly correlated even at large
 $\tau$-separations; the covariance matrix has 
 very small eigenvalues and oscillatory eigenvectors,
 and therefore its inverse is of
 limited practical use in fitting).

%%%%%%%%%%%%%%%%%%%%% TABLE %%%%%%%%%%%%%%%%%%%%%%%%%%%%%%%%%%%%%
%
\begin{table}[t]

\begin{minipage}[c]{5cm}
\small{
\begin{center}
\begin{tabular}{ll}
 \hline \\[-3mm]
 $\tau T$ & 
 $\displaystyle\frac{ Z^{ }_\rmi{pert} G^{ }_\rmi{latt} }{ G^{ }_\rmi{norm}}$ 
 \\[3mm]
 \hline 
          4/48 % =    0.083333
    &  1.8727(43) \\ 
          5/48 % =    0.104167
    &  2.0044(31) \\
          6/48 % =    0.125000
    &  2.1157(25) \\ 
          7/48 % =    0.145833
    &  2.2133(31) \\ 
          8/48 % =    0.166667 
    &  2.3001(44) \\
          9/48 % =    0.187500
    &  2.3793(56) \\
         10/48 % =    0.208333
    &  2.4548(65) \\
 \hline 
\end{tabular} 
\end{center}
}
\end{minipage}%
\begin{minipage}[c]{5cm}
\small{
\begin{center}
\begin{tabular}{ll}
 \hline \\[-3mm]
 $\tau T$ & 
 $\displaystyle\frac{ Z^{ }_\rmi{pert} G^{ }_\rmi{latt} }{ G^{ }_\rmi{norm}}$ 
 \\[3mm]
 \hline 
         11/48 % =    0.229167
    &  2.5300(72) \\
         12/48 % =    0.250000
    &  2.6070(79) \\
         13/48 % =    0.270833
    &  2.6865(86) \\ 
         14/48 % =    0.291667
    &  2.7686(92) \\
         15/48 % =    0.312500
    &  2.8529(96) \\
         16/48 % =    0.333333
    &  2.9391(98) \\
         17/48 % =    0.354167
    &  3.0260(101) \\
 \hline 
\end{tabular} 
\end{center}
} 
\end{minipage}%
\begin{minipage}[c]{5cm}
\small{
\begin{center}
\begin{tabular}{ll}
 \hline \\[-3mm]
 $\tau T$ & 
 $\displaystyle\frac{ Z^{ }_\rmi{pert} G^{ }_\rmi{latt} }{ G^{ }_\rmi{norm}}$ 
 \\[3mm]
 \hline 
         18/48 % =    0.375000
    &  3.1124(107) \\
         19/48 % =    0.395833
    &  3.1962(117) \\ 
         20/48 % =    0.416667
    &  3.2745(130) \\
         21/48 % =    0.437500
    &  3.3441(145) \\
         22/48 % =    0.458333
    &  3.4005(161) \\
         23/48 % =    0.479167
    &  3.4390(176) \\ 
         24/48 % =    0.500000
    &  3.4538(190) \\
 \hline 
\end{tabular} 
\end{center}
} 
\end{minipage}

\vspace*{3mm}

\caption[a]{\small
  The continuum-extrapolated correlator, normalized to 
  \eq\nr{Gnorm}. The errors are statistical and result from linear
  extrapolations in $1/N_\tau^2$ to the continuum limit (cf.\ the text).
 }
\label{table:taudep}
\end{table}
%
%%%%%%%%%%%%%%%%%%%%%%%%%%%%%%%%%%%%%%%%%%%%%%%%%%%%%%%%%%%%%%%%%%%%%

%%%%%%%%%%%%%%%%%%%%%%%%%%%%% SECTION %%%%%%%%%%%%%%%%%%%%%%%%%%%%%%%%%%%%
%
\section{Modelling the spectral function}
\la{se:spectral}

Given the data for the imaginary-time correlator, the next task is to 
constrain the corresponding spectral function. The relation of 
a spectral function $ \rho^{ }_\rmii{E}(\omega) $ to the corresponding
imaginary-time correlator $  G^{ }_\rmii{E}(\tau) $ reads
\be
 G^{ }_\rmii{E}(\tau)
 = 
 \int_0^\infty \! \frac{{\rm d}\omega}{\pi}
 \, \rho^{ }_\rmii{E}(\omega) \, 
 \frac{\cosh [\omega (\frac{\beta }{2}-\tau) ] }
 {\sinh [ \frac{\omega \beta }{2} ] }
 \;.
 \la{relation}
\ee
Even though an inversion of this relation
is possible in principle~\cite{cuniberti} (after
the subtraction of short-distance singularities~\cite{analytic}), the 
problem is ill-posed in practice: large variations of $\rho^{ }_\rmii{E}$
may lead to small changes of $G^{ }_\rmii{E}$. 
Therefore, it is important to constrain the allowed form of 
$\rho^{ }_\rmii{E}$ from general considerations. Here we do 
this by fixing the functional form of $\rho^{ }_\rmii{E}$ at small 
($\omega \ll T$) 
and large frequencies
($\omega \gg T$). 
Subsequently theoretically motivated 
interpolations between the two limits are proposed. 
We consider the spatial volume to be infinite and assume
$\rho^{ }_\rmii{E}$ to be a smooth function of $\omega$, as 
is generally the case in an interacting thermal system.

%%%%%%%%%%%%%%%%%%%%%%%%%%%%% SUBSECTION %%%%%%%%%%%%%%%%%%%%%%%%%%%%%%%%%
%
\subsection{IR and UV asymptotics}

%%%%%%%%%%%%%%%%%%%%%%%%%%%%% SUBSECTION %%%%%%%%%%%%%%%%%%%%%%%%%%%%%%%%%
%
%\subsection{IR regime ($\omega \ll T$)}

In the infrared (IR) regime ($\omega \ll T$), 
the heavy quark momentum diffusion coefficient can  
be defined as~\cite{eucl} 
\be 
 \kappa \equiv \lim_{\omega\to 0} \frac{2 T \rho^{ }_\rmii{E}(\omega)}{\omega}
 \;. \la{icept}
\ee
The approach to this limit appears to be smooth: resummed 
perturbative computations~\cite{eucl}, numerical simulations within
classical lattice gauge theory~\cite{mink}, as well as strong-coupling
computations in analogous theories~\cite{ct,sg2}
suggest that $\rho^{ }_\rmii{E}$
has no transport peak but is rather a monotonically increasing function. 
Therefore, we define the infrared asymptotics through the simplest form
consistent with \eq\nr{icept}: 
\be
   \phi^{ }_\rmii{IR}(\omega) \; \equiv \; 
   \frac{\kappa \omega}{2 T}
 \;. \la{phiIR}   
\ee

%%%%%%%%%%%%%%%%%%%%%%%%%%%%% SUBSECTION %%%%%%%%%%%%%%%%%%%%%%%%%%%%%%%%%
%
%\subsection{UV regime ($\omega \gg T$)}

Consider then the ultraviolet (UV) regime ($\omega \gg T$).
Thanks to asymptotic freedom, the UV behaviour of the spectral function
can be computed in perturbation theory. Denoting by $g^2$
the QCD gauge coupling renormalized in the $\msbar$ scheme, 
and $a^{ }_s \equiv \alpha^{ }_s/\pi \equiv  g^2 / (4\pi^2)$,
the result has the structure
\be
 \rho^{ }_\rmii{E}(\omega) \stackrel{\omega \gg T}{=}
 \Bigl[ \rho^{ }_\rmii{E}(\omega) \Bigr]^{ }_{T = 0} + 
 \rmO\Bigl( \frac{g^4 T^4}{\omega} \Bigr) % + \rmO(g^6)
 \;. \la{UV}
\ee
The leading thermal correction is consistent with the pattern
expected from the Operator Product Expansion~\cite{sch}
(for $\Nf=0$ the coefficient of this correction is negative~\cite{rhoE}). 
The vacuum part has the form 
\ba
 \Bigl[ \rho^{ }_\rmii{E}(\omega) \Bigr]^{ }_{T = 0} 
 & = &  \frac{g^2 \CF\, \omega^3}{6\pi}
 \, \Bigl[
   r^{ }_{10} + (r^{ }_{20} + r^{ }_{21} \ell)\, a^{ }_s 
  + (r^{ }_{30} + r^{ }_{31} \ell + r^{ }_{32}\ell^2)\, a^{2}_s + \rmO(a_s^3)
  \Bigr]
 \;, \hspace*{8mm} \la{structure}
\ea
where $\CF \equiv (\Nc^2 - 1)/(2\Nc) = 4/3$ and 
$\ell \equiv \ln (\bmu^2 / \omega^2)$, with $\bmu$ denoting
the renormalization scale. The three first coefficients read~\cite{rhoE}
\ba
 r^{ }_{10} & = & 1 \;, \\ 
 r^{ }_{20} & = &  
 \Nc \, 
 \biggl( 
   \frac{149}{36}-\frac{11\ln2}{6}-\frac{2\pi^2}{3}
 \biggr) 
 - \Nf\,
 \biggl(
  \fr59 - \frac{\ln 2}{3} 
 \biggr)
 \;, \la{r20} \\ 
 r^{ }_{21} & = & \frac{11\Nc - 2\Nf}{12} \;, 
\ea
where $\Nf$ denotes the number of light dynamical quarks
($\Nf = 0$ in our study). 
The coefficient $r^{ }_{30}$ and higher-order terms
are presently unknown. However, the general 
structure of \eq\nr{structure}, together with the knowledge
that $\rho^{ }_\rmii{E}$ requires no renormalization 
in dimensional regularization~\cite{eucl}, 
is sufficient for determining the asymptotics of $\rho^{ }_\rmii{E}$. 

Indeed, suppose that we choose 
the renormalization scale as
$\bmu = \omega$ for $\omega \gg \Lambdamsbar$.
Then $\ell =0$, 
$a^{ }_s \sim \ln^{-1}(\omega/\Lambdamsbar)$, 
and we get
\be
 \Bigl[ \rho^{ }_\rmii{E}(\omega) \Bigr]^{ }_{T = 0} 
 \; \stackrel{\omega \gg T}{=} \; 
 \phi^{(a)}_\rmii{UV}(\omega) \, \biggl[  
   1 + \rmO\biggl( \frac{1}{\ln(\omega/\Lambdamsbar)} \biggr)
 \biggr]
 \;, 
\ee
where we have defined 
\be
  \phi^{{(a)}}_\rmii{UV}(\omega) \; \equiv \;
  \frac{g^2(\bmu^{ }_\omega) \CF\, \omega^3}{6\pi}
 \;, \quad
 \bmu^{ }_\omega \equiv \mbox{max}(\omega,\pi T)
 \;. \la{phiUV}
\ee
Formally, we can reduce the correction to be of quadratic rather 
than linear order 
in $\ln^{-1} (\omega/\Lambdamsbar)$ by including $r^{ }_{20}$ 
in the asymptotics, so let
\be
  \phi^{{(b)}}_\rmii{UV}(\omega) \; \equiv \;
  \phi^{{(a)}}_\rmii{UV}(\omega) \, 
  \Bigl[ 
    1 + (r^{ }_{20} + r^{ }_{21} \ell) \, a^{ }_s (\bmu^{ }_\omega)
  \Bigr]
  \;. \la{phiUV2}
\ee
However, since the convergence of an expansion proceeding in inverse 
logarithms could be slow, and since our knowledge of 
$\omega/\Lambdamsbar = (\omega/T)\times (T/\Lambdamsbar)$ 
is imperfect due to uncertainties in $T/\Lambdamsbar$~\cite{betac}, 
we treat \eqs\nr{phiUV} and \nr{phiUV2} on an equal footing in the following. 
For evaluating $g^2(\bmu^{ }_\omega)$ and $a^{ }_s(\bmu^{ }_\omega)$, 
we have used 4-loop running~\cite{beta}.

%%%%%%%%%%%%%%%%%%%%%%%%%%%%% SUBSECTION %%%%%%%%%%%%%%%%%%%%%%%%%%%%%%%%%
%
\subsection{Interpolations}
\la{ss:models}

Equations \nr{phiIR} and \nr{phiUV} determine the limiting behaviours 
of the spectral function. For $\omega \ll T$, 
given the flatness observed in non-perturbative simulations~\cite{mink}, 
we expect corrections to $\phi^{ }_\rmii{IR}$ to be given by 
a convergent power series in $\omega$. 
For $\omega \gg T$ the corrections are suppressed by inverse logarithms
of $\omega / T$, and it 
is important to account for these corrections. 
In order to incorporate the different types of corrections in the 
two regimes, we map the interval $\omega \in (0,\infty)$ to the
interval $(0,1)$ by introducing 
\be
 x \; \equiv \; \ln\Bigl( 1 + \frac{\omega}{\pi T} \Bigr) \in (0,\infty) 
 \;, \quad
 y \; \equiv \; \frac{x}{1+x} \in (0, 1)
 \;. 
\ee
For small $y$ we have 
\be
 y \approx x \approx \frac{\omega}{\pi T} \;, \quad \omega \ll T 
 \;, \la{y_small}
\ee
whereas around the other end we get
\be
 1-y \approx \frac{1}{x} \approx \frac{1}{\ln[\omega/(\pi T)]}
 \;, \quad \omega \gg T
 \;. \la{y_large}
\ee 
So, we can capture corrections to the asymptotics by a function 
in $y$ which vanishes at $y = 0$ and $y=1$ and is a polynomial
or a power series in between. 
A convenient choice is to employ 
trigonometric functions, 
\be
 e^{(\alpha)}_n(y) \equiv \sin(\pi n y)
 \;. \la{e_alpha}
\ee
Another possibility, in principle preferable if we know that corrections
are only quadratic in the variables in \eqs\nr{y_small}, \nr{y_large}
(cf.\ the discussion preceding \eq\nr{phiUV2}), is 
\be
 e^{(\beta)}_n(y) \equiv \sin(\pi y) \sin(\pi n y)
 \;. \la{e_beta}
\ee
With these bases we are led to define general models
($
 \mu \in \{ \alpha,\beta \} 
$,
$ 
 i \in \{ a,b\}
$), 
\ba
 \mbox{model 1}: & & 
 \rho^{(1\mu{}i)}_\rmii{E}(\omega) \; \equiv \; 
 \Bigl[ 1 + \sum_{n=1}^{n_\rmii{max}} c^{ }_n e^{(\mu)}_n(y) \Bigr]
 \Bigl[ \phi^{ }_\rmii{IR}(\omega) + \phi^{(i)}_\rmii{UV}(\omega) \Bigr]
 \;, \la{model1} \\ 
 \mbox{model 2}: & & 
 \rho^{(2\mu{}i)}_\rmii{E}(\omega) \; \equiv \; 
 \Bigl[ 1 + \sum_{n=1}^{n_\rmii{max}} c^{ }_n e^{(\mu)}_n(y) \Bigr]
 \sqrt{ \bigl[ \phi^{ }_\rmii{IR}(\omega)\bigr]^2_{ } + 
        \bigl[ \phi^{(i)}_\rmii{UV}(\omega)\bigr]^2_{ } }
 \;. \la{model2}
\ea
The difference between these interpolations is that the latter imposes
a more rapid crossover from the IR to the UV asymptotics, for instance
for large $\omega$ we have 
$
  \sqrt{ \phi^{ 2}_\rmii{IR} + \phi^{2 }_\rmii{UV} } 
 \approx  
  \phi^{ }_\rmii{UV} + \frac{ \phi^{ 2}_\rmii{IR} }
                            { 2 \phi^{ }_\rmii{UV}}
 \sim \omega^3 + T^4/\omega   
$
in qualitative accordance with the  functional form 
expected from \eq\nr{UV}.\footnote{%
 Note that, for $n_\rmii{max}\to\infty$, such a construction can be viewed
 as a general parametrization of an arbitrary spectral function. We call 
 our parametrizations ``models'' because the problems discussed in 
 \se\ref{ss:fits} necessitate keeping $n_\rmii{max}$ relatively small 
 and stabilizing the subsequent fits through additional input.    
 } 
Finally, we also consider a simple 2-parameter ansatz separating the 
IR and UV regimes completely: 
\ba
 \mbox{model 3}: & & 
 \rho^{(3i)}_\rmii{E}(\omega) \; \equiv \; 
 \mathop{\mbox{max}}
 \Bigl[  {\phi}^{ }_\rmii{IR}(\omega) , 
    c\, \phi^{(i)}_\rmii{UV}(\omega) 
 \Bigr]
 \;, \hspace*{8mm} \la{model3}
\ea
where $c$ is treated as a free parameter, reflecting 
uncertainties in the renormalization factor in \eq\nr{ZE}
(in practice we find $c=1.05(1)$ from the fit for $i=a$). 

%%%%%%%%%%%%%%%%%%%%%%%%%%%%% SUBSECTION %%%%%%%%%%%%%%%%%%%%%%%%%%%%%%%%%
%
\subsection{Fitting strategy}
\la{ss:fits}

The inversion of \eq\nr{relation} represents an ill-posed problem, 
so it should not come as a surprise that it is, in general, not possible to 
find a ``stable'' fit describing the data. In other words, the 
$\chi^2$-function may possess an extremely shallow minimum, or multiple
minima. Empirically we find, however, that these ambiguities 
are largely related to the UV behaviour of the spectral function. 
With some further input, the UV behaviour can be stabilized, yet
the IR coefficient $\kappa$ in which we are mostly interested,
remains fairly unaffected. 
% This is comforting also in view of the 
% fact that  our study relies on a perturbative
% renormalization factor, \eq\nr{ZE}, 
% which induces its own uncertainty.

Concretely, we have implemented two strategies for the fitting, which 
incorporate an implicit or explicit stabilization of the UV contribution.
After defining\footnote{% 
  We have also carried out tests with the covariance matrix, but in its
  full form this does not yield sensible results as alluded to above. 
  If a sufficient infrared cutoff is imposed on the smallest eigenvalues, 
  the results are consistent within errors with the procedure described here.} 
\be
 \chi^2 \; \equiv \; 
 \sum^{ }_{\tau_i} \biggl[ \frac{ G^{ }_\rmi{meas}(\tau_i) - 
   G^{ }_\rmi{model}(\tau_i) }{\delta G^{ }_\rmi{meas}(\tau_i)} \biggr]^2  
 \;, \la{chi2}
\ee
where $\delta G^{ }_\rmi{meas}$ denote the errors from 
table~\ref{table:taudep}, the following alternatives 
are considered:

\begin{itemize}

\item[(i)]
In the first strategy, we search for a minimum of $\chi^2$ for
a result based on \eq\nr{relation}, taking a jackknife
sample as the measured result. The search is performed using
the algorithm described in ref.~\cite{search2}, 
taking $\kappa/T^3 = 1.0$, $c_n =0.0$ as a starting point. The 
search is stopped after 200 iterations, by which time an excellent
representation of the data has in general been reached
(with $\chi^2$/d.o.f.\ $\sim 0.3 - 0.5$ for a typical sample from 
the jackknife ensemble), with coefficients $|c^{ }_n| \lsim 0.2$.

\item[(ii)]
In the second strategy, we stabilize the fit 
by imposing the constraint
\be
 \rho^{(n\mu{}i)}_\rmii{E}(\omega^{ }_\rmi{max}) \; \equiv \; 
 \phi^{(i)}_\rmii{UV}(\omega^{ }_\rmi{max})
 \;, \la{constraint}
\ee
where we choose $\omega^{ }_\rmi{max}\equiv 1000\, T$. This constraint
imposes a relation between $\kappa$ and $c^{ }_n$, so that there are
only $n^{ }_\rmi{max}$ free parameters. We find that with this 
constraint, a unique minimum of $\chi^2$ can always be found, 
with $\chi^2$/d.o.f.\ $\sim 0.2 - 0.6$. 
The practical search was performed with the routine
described in ref.~\cite{search1}. Because of the artificial
constraint at large $\omega/T$, the fit is carried out only
to distances $\tau T \ge 6/48$. %  (cf.\ table~\ref{table:taudep}).

\end{itemize} 

\noindent
In the following we describe results from both procedures, demonstrating
that they yield similar results for $\kappa$, particularly for our 
preferred ``model~2'' (cf.\ \eq\nr{model2}).

In the case of the 2-parameter ``model 3'' defined in \eq\nr{model3}, a global
minimum of $\chi^2$ is readily found without further input. The price
to pay for this stability is that the model does not describe the data
particularly well, having $\chi^2/\mbox{d.o.f.} \approx 17.5$ for
$\phi^{(a)}_\rmii{UV}$ (we do not show results for
$\phi^{(b)}_\rmii{UV}$ for which $\chi^2/\mbox{d.o.f.} \approx 68$).  

In addition to these fits, we have also made use of a variant of the  
Backus-Gilbert method (BGM)~\cite{bgm1,bgm2} (see refs.~\cite{bgm3,bgm4}
for previous applications in lattice QCD). The goal of this method is
not to reconstruct the spectral function itself, but rather an averaged
version thereof. With very precise data, the averaging kernel could be made
optimally narrow in a certain sense. In practice, the finite precision
of the data necessitates a regularization of an ill-defined matrix inversion; 
this is characterized by a parameter $\lambda$ for which we 
use $\lambda = 10^{-4}$ rather than the theoretically optimal $\lambda = 1$
(cf.\ ref.~\cite{bgm4}). 
As a consequence, our estimate of the infrared limit of the spectral
function amounts to a weighted average over the range 
$0 \le \omega \lsim 10 T$. Fortunately, if there is little structure
in this range, the result should be reasonable. 
Moreover it is possible to rescale the kernel in order to further
remove known structures; we insert 
$\rho^{ }_\rmii{E}(\omega) = 
 [\rho^{ }_\rmii{E}(\omega) / \phi(\omega) ]\,\phi(\omega)$ in \eq\nr{relation}
with $\phi(\omega) \equiv (\omega \beta)^3 / \tanh^2(\omega\beta/4)$, and
subsequently include $\phi$ in the kernel.  
To keep the matrix size manageable, 
only $\tau / \beta = (4 - 21)/48$ were used for this analysis. 
If the final result for the averaged $\rho^{ }_\rmii{E}(\omega)$ is 
inserted back into \eq\nr{relation}, it yields a representation
of the original data with $\chi^2/\mbox{d.o.f.} \approx 41$. 

 We end by remarking that apart from these novel approaches, we have 
 also applied the standard Maximum Entropy Method (MEM) to our problem. 
 MEM requires the specification of a default model as input. If we
 take as the default model our model 1 or 2, which yield
 a Euclidean correlator agreeing with lattice data everywhere (within
 statistical errors), then MEM reproduces the default model as its 
 output (within statistical errors). In other words, MEM does not
 help to constrain the result beyond our analysis. 

%%%%%%%%%%%%%%%%%%%%%%%%%%%%%%%%% FIGURE %%%%%%%%%%%%%%%%%%%%%%%%%%%%%%%%%
\begin{figure}[t]

%\vspace*{-3cm}

\centerline{%
 \epsfysize=7.5cm\epsfbox{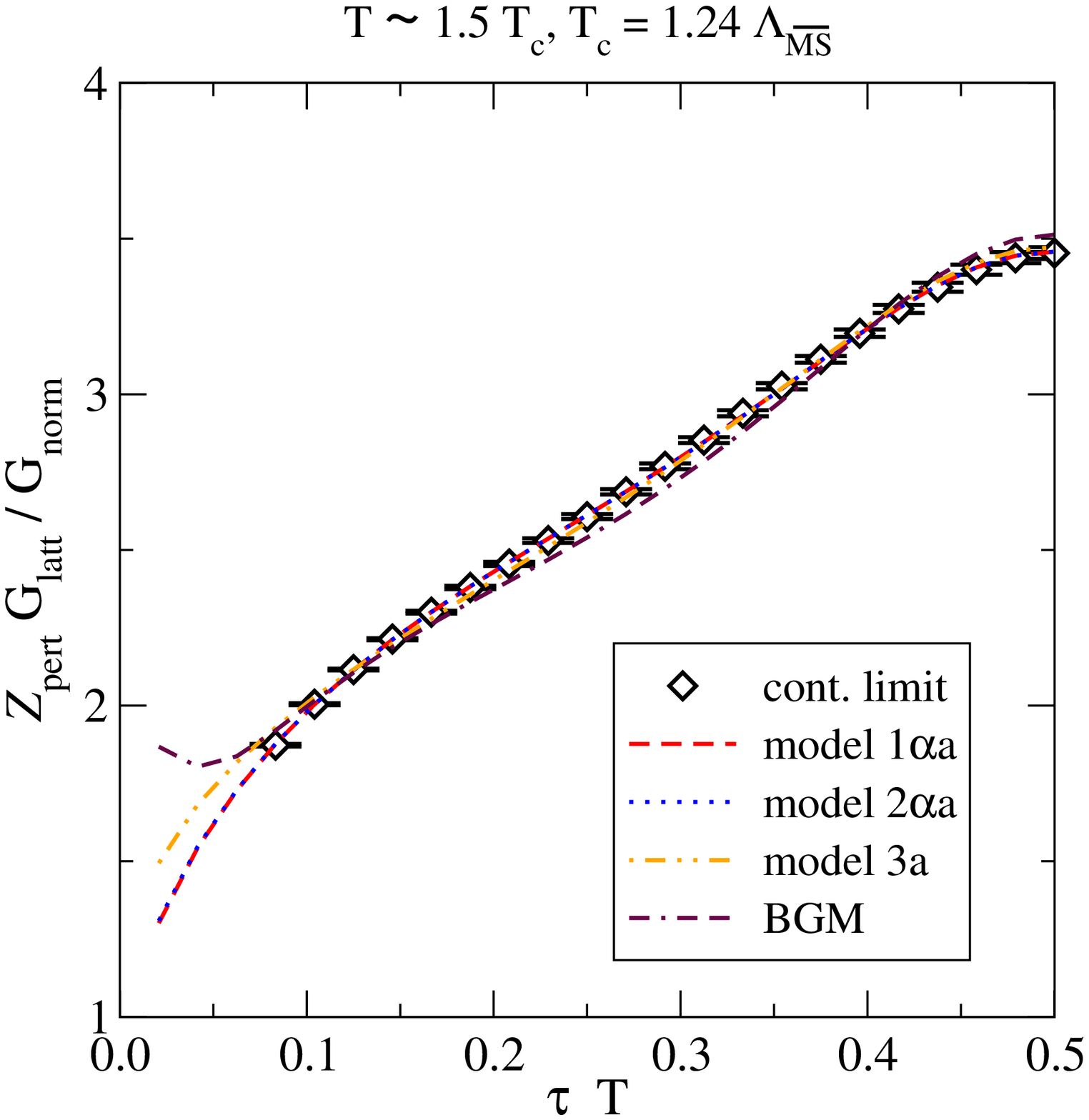}%
~~~\epsfysize=7.5cm\epsfbox{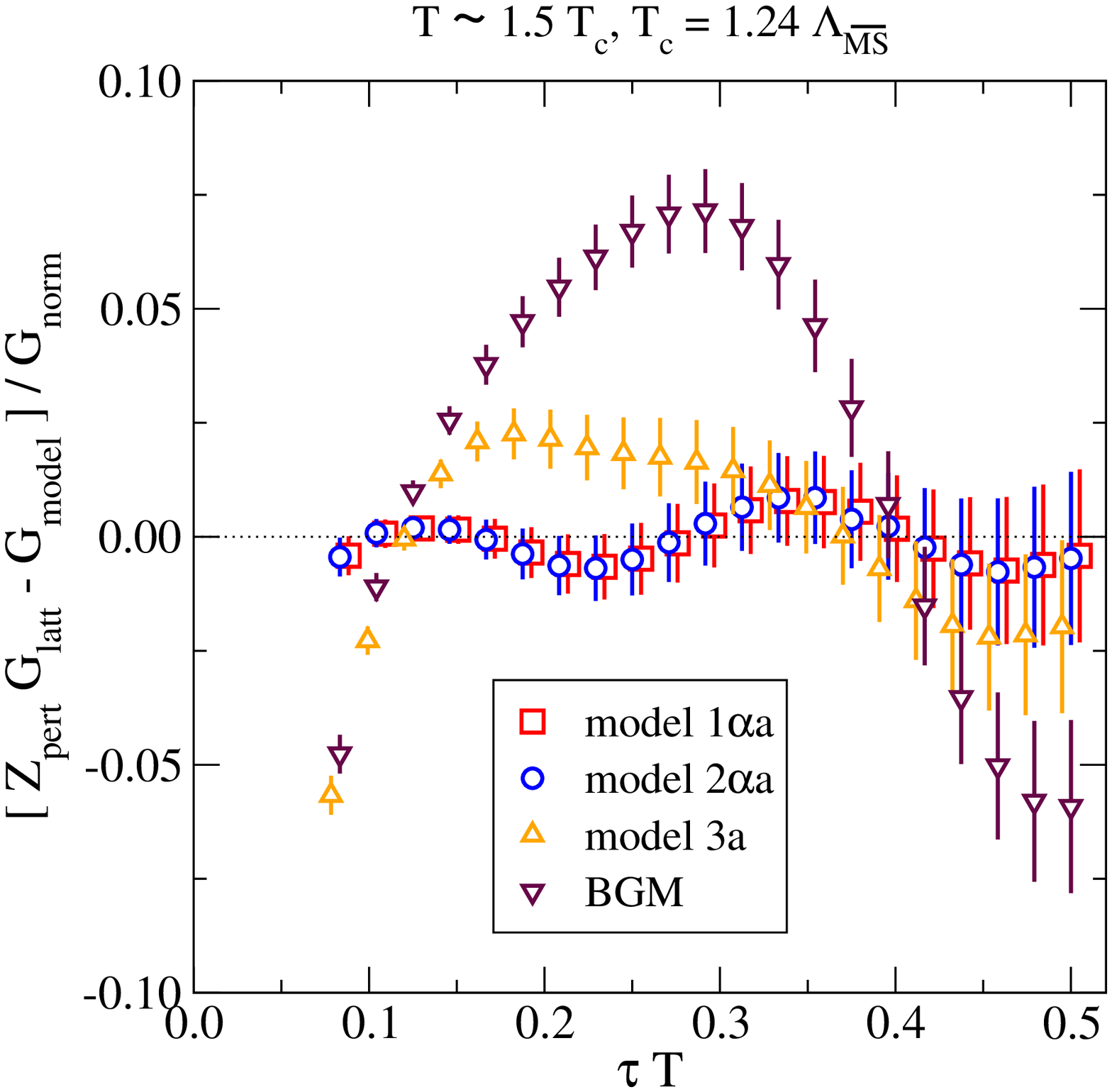}
}

\caption[a]{\small
 Left: Continuum-extrapolated lattice data
 (cf.\ table~\ref{table:taudep}) and examples
 of model results, as described
 in the text. The results have been normalized to \eq\nr{Gnorm}. 
 Right: The differences between the data and the models. 
 For better visibility, the data points corresponding to 
 models $1{\alpha}a$ and $3a$
 have been displaced slightly. 
}

\la{fig:taudep}
\end{figure}
%%%%%%%%%%%%%%%%%%%%%%%%%%%%%%%%%%%%%%%%%%%%%%%%%%%%%%%%%%%%%%%%%%%%%%%%%%%

%%%%%%%%%%%%%%%%%%%%%%%%%%%%%%%%% FIGURE %%%%%%%%%%%%%%%%%%%%%%%%%%%%%%%%%
\begin{figure}[t]

%\vspace*{-3cm}

\centerline{%
 \epsfysize=7.5cm\epsfbox{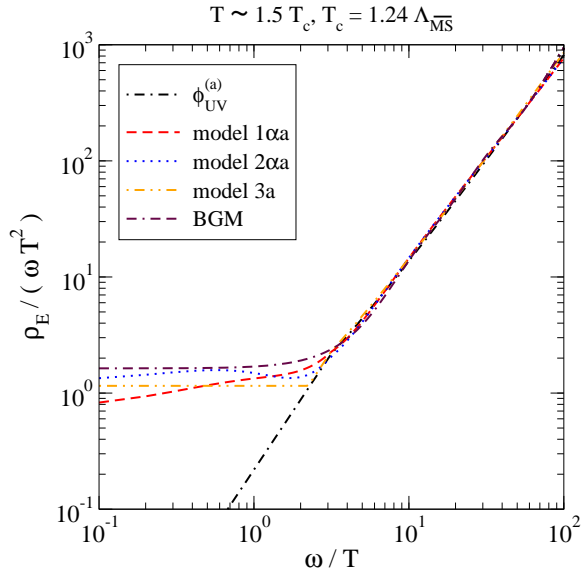}%
% ~~~\epsfysize=7.5cm\epsfbox{taudep_diff.eps}
}

\caption[a]{\small
 Examples of model spectral functions, compared 
 with $\phi^{(a)}_\rmii{UV}$ from \eq\nr{phiUV}. In the BGM case the 
 curve shown has been obtained by replacing the full covariance matrix with
 a diagonal one possessing the errors shown in table~\ref{table:taudep}, 
 but we have obtained BGM results with full and IR-regularized covariance
 matrices as well; they are included in the average shown 
 in \fig\ref{fig:kappa}. 
}

\la{fig:rho}
\end{figure}
%%%%%%%%%%%%%%%%%%%%%%%%%%%%%%%%%%%%%%%%%%%%%%%%%%%%%%%%%%%%%%%%%%%%%%%%%%%

%%%%%%%%%%%%%%%%%%%%%%%%%%%%% SUBSECTION %%%%%%%%%%%%%%%%%%%%%%%%%%%%%%%%%
%
\subsection{Estimation of $\kappa$}
\la{ss:results}

%%%%%%%%%%%%%%%%%%%%%%%%%%%%%%%%% FIGURE %%%%%%%%%%%%%%%%%%%%%%%%%%%%%%%%%
\begin{figure}[t]

%\vspace*{-3cm}

\centerline{%
 \epsfysize=7.5cm\epsfbox{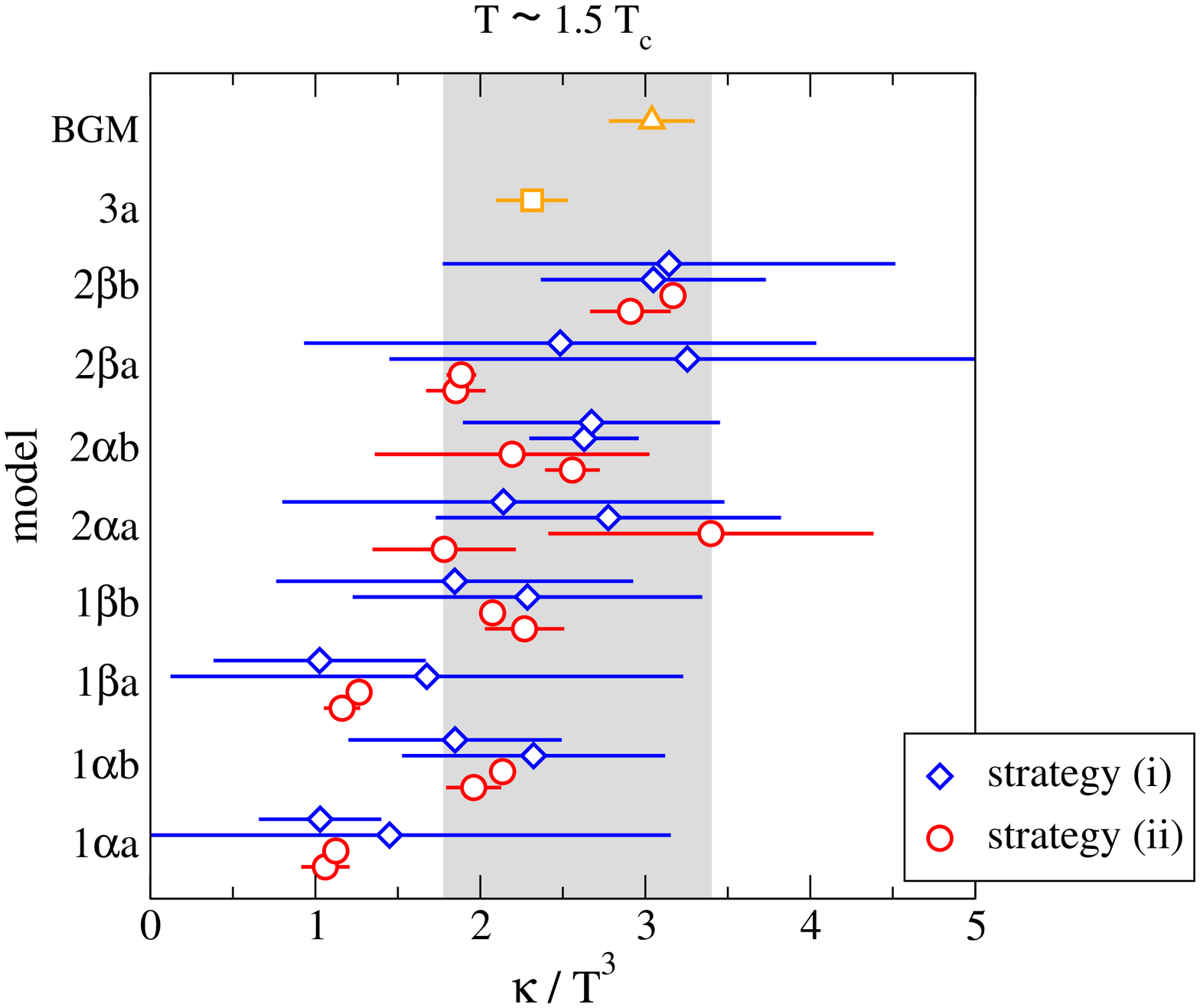}%
% ~~~\epsfysize=7.5cm\epsfbox{taudep_diff.eps}
}

\caption[a]{\small
 Fit results based on different models (cf.\ \se\ref{ss:models}) and
 different fitting strategies (cf.\ \se\ref{ss:fits}). In each case, 
 the lower data point corresponds to $n^{ }_\rmi{max} = 4$, 
 the higher to $n^{ }_\rmi{max} = 5$. For model $3a$ (cf.\ \eq\nr{model3})
 and BGM (cf.\ \se\ref{ss:fits}) systematic errors are  
 larger than those shown. The grey band illustrates our final
 estimate, given in \eq\nr{kappa_final} and based on models 2, $3a$ and BGM.
}

\la{fig:kappa}
\end{figure}
%%%%%%%%%%%%%%%%%%%%%%%%%%%%%%%%%%%%%%%%%%%%%%%%%%%%%%%%%%%%%%%%%%%%%%%%%%%

We now turn to estimating $\kappa$.
In \se\ref{ss:models} we have introduced a large 
class of fit functions: 8 separate models 
$\rho^{(n\mu{}i)}_\rmii{E}$, 
$
 n \in \{1,2\}
$,
$
 \mu \in \{ \alpha,\beta \} 
$,
$ 
 i \in \{ a,b\}
$, 
each of which depends on a parameter $n^{ }_\rmi{max}$. 
In addition we have introduced 
two fitting strategies (cf.\ \se\ref{ss:fits}). 
In the following, we choose $n^{ }_\rmi{max} = 4,5$ and
demonstrate that, for a given model, both strategies
yield similar results within error bars. (We have also used 
$n^{ }_\rmi{max} = 3$ and $n^{ }_\rmi{max} = 6$; 
these yield nothing qualitatively new.)
The agreement is more significant for our preferred model~2. 
The spread between the models is
interpreted as an indication of the systematic uncertainty
of our determination. Model $3a$ and the BGM approach are
included as further crosschecks; in the BGM case the spread of 
results originating from different
versions of covariance matrices (cf.\ caption of \fig\ref{fig:rho}) 
is included in the error shown. 

The fits are carried out to samples from a jackknife ensemble, 
and the ensemble is used for determining statistical errors. 
Results from the different models are illustrated in 
\figs\ref{fig:taudep}, \ref{fig:rho} and \ref{fig:kappa}
(the fits in \figs\ref{fig:taudep} and \ref{fig:rho} are based on
strategy (i)). 
Given that model~2 yields more stable
results within the other variations, and that it is theoretically
better justified than model 1
(cf.\ the discussion below \eq\nr{model2}),
we make use of it
in the following. Based on the central values from
\fig\ref{fig:kappa} for models 2, $3a$ and BGM we estimate 
\be
  \kappa / T^3 = 1.8 \ldots 3.4
  \;. \la{kappa_final}
\ee  
This result is illustrated in \fig\ref{fig:kappa} with a grey band. 
Given that the result is dominated by systematic uncertainties which
may be asymmetric but are essentially impossible to estimate, we 
refrain from citing a central value here.\footnote{%
 It is remarkable that our result is consistent with previous ones
 obtained at a single lattice spacing and with very rough modelling
 of the spectral shape~\cite{kappaE,latt_c}. 
 However, the uncertainties related
 to the continuum extrapolation and to analytic continuation are
 now much closer to being under control. 
 } 

%%%%%%%%%%%%%%%%%%%%%%%%%%%%% SECTION %%%%%%%%%%%%%%%%%%%%%%%%%%%%%%%%%%%%
%
\section{Conclusions}
\la{se:concl}

The purpose of this paper has been to estimate the heavy-quark
momentum diffusion coefficient, defined 
through \eqs\nr{GE_final}, \nr{relation} and \nr{icept}.
Compared with previous works~\cite{latt_a,kappaE,latt_c}, 
we have carried out a continuum extrapolation of the 
imaginary-time correlator (cf.\ \se\ref{se:continuum}) and
discussed the systematics related to estimating the 
corresponding spectral function (cf.\ \se\ref{se:spectral}).
The final result of our analysis is given in \eq\nr{kappa_final}. 
 It is remarkable that, despite the ill-posed nature of analytic 
 continuation, our novel approach permits us to obtain relatively
 stable results, and to strongly constrain the order of magnitude of 
 the heavy quark momentum diffusion coefficient in the continuum limit.

In the non-relativistic limit (i.e. for a heavy quark mass $M\gg \pi T$)
$\kappa$ is related to the diffusion coefficient $D$ as $D=2 T^2 / \kappa$,
and to the drag coefficient as $\eta^{ }_\rmii{$D$} = \kappa / (2 M T)$, 
where $M$ is a heavy quark kinetic mass~\cite{eucl}. 
The drag coefficient can also be interpreted as the kinetic equilibration
time scale associated with heavy quarks: 
$\tau^{ }_\rmi{kin} = \eta^{-1}_\rmii{$D$}$.
For a conversion to physical units, we use 
$\rO\Tc = 0.7457(45)$~\cite{betac} 
and $\rO  = 0.47(1)$~fm~\cite{scale}.
With these conversions, our result $\kappa / T^3 = 1.8 - 3.4 $ yields  
an estimate for the time 
scale associated with the kinetic equilibration of heavy quarks,
\be
 \tau^{ }_\rmi{kin} = 
 \frac{1}{\eta^{ }_\rmii{$D$}} = 
 (1.8 \ldots 3.4)\, \biggl(\frac{\Tc}{T}\biggr)^2  %% 6/x 
 \biggl(\frac{M}{\mbox{1.5~GeV}}\biggr)\,\mbox{fm/c}
 \;.
\ee
Close to $\Tc$, charm quark kinetic equilibration appears therefore to be 
almost as fast as that of light partons, for which a time scale
$\sim 1$~fm/c is generally considered appropriate. 
For the diffusion coefficient we obtain
\be
 D\, T = 0.59 \ldots 1.1    %% \frac{2}{x}
 \;. 
\ee
This can be compared with the values $D T\, \gsim \, 0.13$ obtained
for light quarks in quenched QCD in the continuum limit~\cite{cond3,cond4}.
It would be ``natural'' for the $D$ of heavy quarks to be of the same
order of magnitude but somewhat larger than that of light quarks, given that 
heavy quarks should feel slightly weaker interactions. 

It is also interesting to compare our result for $\kappa$
with an NLO perturbative
computation~\cite{chm1}. That result is of the form 
$\kappa / T^3  =  \alphas^2\, (c^{ }_1 + c^{ }_2\, \alphas^{1/2})$, where
$c^{ }_1, c^{ }_2$ are coefficients given in ref.~\cite{chm1}
($c^{ }_1$ involves a logarithmic dependence on $\alphas$).
In the absence of corrections of relative order $\alphas$, it is 
not possible to estimate the renormalization scale at which 
$\alphas$ should be evaluated. Nevertheless, the 
result shown in fig.~3 of ref.~\cite{chm1}  agrees with our 
\eq\nr{kappa_final} if we set $\alphas = 0.20 - 0.26$, which
is in full accordance with the range generally used in 
heavy ion collision phenomenology. 

Many possible directions can be envisaged for future investigations. 
Improved statistical precision is crucial for moving towards
model-independent analytic continuation~\cite{analytic}.
Other temperatures than just $T \sim 1.5\Tc$ should be considered. 
The determination of the renormalization factor 
$\mathcal{Z}^{ }_\rmii{E}$ in \eq\nr{renorm} should be promoted to
the non-perturbative level. 
It would be important to understand whether the
heavy-mass limit is justified for charm quarks (or only for
bottom quarks); this can in principle be studied by using
the full relativistic formulation for measuring 
current-current correlation functions~\cite{rel,ohno_2}, 
even though then the structure of the spectral function
is more complicated and analytic continuation is 
even more difficult to get under reasonable control.  
Finally, estimating effects from 
dynamical quarks is important for phenomenological applications. 

%%%%%%%%%%%%%%%%%%%%%%%%%%%%% SECTION %%%%%%%%%%%%%%%%%%%%%%%%%%%%%%%%%%%%
%
\section*{Acknowledgments}                                    
                       
We thank J.~Langelage and 
M.~M\"uller for collaboration at initial stages of this project, and
H.~Sandmeyer for his work on the $80^3\times 20$ ensemble. 
M.L is grateful to H.B.~Meyer for helpful discussions.  
Our work has been supported
in part by the DFG under grant GRK881, 
by the SNF under grant 200020-155935,
by the European Union through HadronPhysics3
(grant 283286) and ITN STRONGnet
(grant 238353), and by the V\"ais\"al\"a Foundation.
Simulations were
performed using JARA-HPC resources at the RWTH Aachen and JSC J\"ulich
(projects JARA0039 and JARA0108), JUDGE/JUROPA at the JSC J\"ulich, 
the OCuLUS Cluster at the Paderborn Center for Parallel
Computing, and the Bielefeld GPU cluster.

%%%%%%%%%%%%%%%%%%%%%%%%%%%%%%%%%%%%%%%%%%%%%%%%%%%%%%%%%%%%%%%%%%%%%%%%%%%
%

\end{document}